\RequirePackage{fix-cm}
\documentclass[smallextended]{svjour3}       
\smartqed  
\usepackage{graphicx}
\usepackage{braket}
\usepackage{amsmath,amssymb}
\usepackage[utf8]{inputenc}

 \journalname{Journal of Statistical Physics}
\begin{document}

\title{
Quantum Counterpart of Classical Equipartition of Energy
}


\author{ Jerzy Łuczka}


\institute{                   \email{jerzy.luczka@us.edu.pl}   \\
          Institute of Physics,  University of Silesia, 41-500 Chorz\'{o}w, Poland \\
              Institute of Physics, University of Augsburg, 86135 Augsburg, Germany}

\date{Received: date / Accepted: date}

\maketitle

\begin{abstract}
It is shown that the recently proposed quantum analogue of classical energy equipartition theorem for two paradigmatic, exactly solved models (i.e., a free Brownian particle and a dissipative harmonic oscillator) also holds true for all quantum systems  which are composed of an arbitrary number of  non-interacting or interacting particles,  subjected to  any confining potentials and coupled to thermostat with arbitrary coupling strength.
 
\end{abstract}


\section{Introduction}
\label{intro}

In classical statistical physics, the theorem on equipartition of kinetic energy  is one of the most universal relation \cite{huang,terlecki}. It states  that for a system in  thermodynamic equilibrium of temperature $T$, the mean kinetic energy $E_k$ per one degree of freedom is equal to $E_k=k_B T/2$, where $k_B$ is the Boltzmann constant 
\cite{waterston,boltzmann}. It does not depend on a number of particles in the system, the form of the  potential  force which acts on them, the form of interaction between particles and   strength of coupling between the system and thermostat. It depends only on the thermostat temperature $T$.  On the contrary, for quantum systems, the mean kinetic energy is not equally shared among all degrees of freedom and the theorem fails. The quite natural  question arises whether one can formulate a similar and  universal relation for the mean kinetic energy of quantum systems at the thermodynamic equilibrium state. Recently, in a series of papers \cite{PRA,SR,JPA}, the authors have proposed quantum analogue of the classical energy equipartition theorem. For a system of one degree of freedom this quantum counterpart, which is called the  energy {\it partition} theorem, has the following appealing form: 
\begin{equation}\label{Ek0}
E_k  =  \int_0^{\infty}  \mathcal{E}_k(\omega)\mathbb{P}(\omega)\, d\omega,  
\end{equation}
where 
\begin{equation}\label{ho1}
\mathcal{E}_k(\omega) = 
\frac{\hbar \omega}{4} \coth\left[{\frac{\hbar \omega}{ 2k_BT}}\right]
\end{equation} 
has the same form as thermally  averaged  kinetic energy of the harmonic oscillator with the frequency $\omega$  weakly coupled to thermostat of temperature $T$ \cite{feynman}.   
The function $\mathbb{P}(\omega)$ has all properties of  a probability density on a positive half-line of real numbers meaning that
\begin{eqnarray}
	\mathbb{P}(\omega) \ge 0, \label{poz}\\
	\int_0^{\infty} {\mathbb P}(\omega)\, d\omega = 1. \label{nor}
\end{eqnarray}
The explicit form of $\mathbb{P}(\omega)$ has been derived for two exactly solved quantum systems: a free Brownian particle \cite{PRA} and a dissipative harmonic oscillator \cite{SR}. In these papers \cite{PRA,SR}, thermostat is composed of quantum harmonic oscillators ({\`a} la Caldeira-Leggett \cite{zwanzig,caldeira,ph}) and the above interpretation of $\mathcal{E}_k(\omega)$ as their  mean kinetic energy per one degree of freedom is fully justified. Because ${\mathbb P}(\omega)$ is a probability density, Eq. (\ref{Ek0}) can be rewritten in the form 
\begin{equation}\label{Ek1}
E_k  =  \langle \mathcal{E}_k \rangle,   
\end{equation}
where $\langle \mathcal{E}_k \rangle$ is a mean value of the function $\mathcal{E}_k(\xi)$ of some  random variable $\xi$ distributed according to the probability density ${\mathbb P}$. In the Caldeira-Leggett model, $\xi$ can be interpreted as a random frequency of  harmonic oscillators forming the thermostat which should be infinitely extended, i.e. the thermodynamic limit for the thermostat should be carried out in order to guarantee a continuous spectrum of the thermostat oscillators frequencies.



\section{Universal Relation for Kinetic Energy of Quantum Systems}

Here, we want to prove a relation similar to Eq. (\ref{Ek0}) for a class of quantum systems for which the concept of kinetic energy has sense (e.g spin systems are outside of this class).  More precisely, we study a quantum  system $S$ coupled to a heat bath (thermostat, environment) $B$. The  composite system $S+B$  is in a  Gibbs  equilibrium state of temperature $T$  defined  by the density operator 
\begin{equation} \label{rho}
\rho =Z^{-1} \, e^{-H/k_BT}, \quad Z = \mbox{Tr}\left[ e^{-H/k_BT}\right]
\end{equation}
 and 
\begin{equation} \label{H}
H =  H_S + H_{int} + H_B 
\end{equation}
is the Hamiltonian of the composite system $S+B$. Next, 
    \begin{equation} \label{HS}
    H_S = \sum_j \frac{p^2_j}{2M_j} + \sum_j U_S(x_j) + \sum_{j,k} V_S(x_j, x_k)
    \end{equation}
      is the Hamiltonian of the system $S$ and  
        \begin{equation} \label{HSE}
    H_{int} = \sum_{j,n} \lambda_{jn} \;V(x_j, X_n) 
    \end{equation}
    is the Hamiltonian of interaction of the system $S$ with the thermostat $B$.  Finally, $H_B$ is the Hamiltonian of thermostat $B$. Its explicit form is now not relevant. The set of parameters $\{\lambda_{jn}\}$ characterizes the coupling strength. The coordinate and momentum operators $\{x_j, p_j\}$ refer to the system $S$ and the operators $\{X_n\}$ refer to the  thermostat $B$. All coordinate and momentum operators obey canonical equal-time commutation relations. We assume that  all components of the Hamiltonian  (\ref{H}) fulfil required conditions to ensure a well defined thermodynamic equilibrium state of the composite system $S+B$ in the thermodynamic limit for the thermostat.  \\

{\bf Theorem 1} {\it  The mean  kinetic energy  per one degree of freedom of the system $S$  can be  expressed in a universal form as  } 
\begin{equation}\label{Ekj}
E_k^{(j)}  =  \langle \mathcal{E}_k \rangle^{(j)} = \int_0^{\infty}  \mathcal{E}_k(\omega)\mathbb{P}_j(\omega)\,  d\omega,    
\end{equation}
{\it where}
\begin{equation}  \label{Ej}
E_k^{(j)} = \langle \frac{p_j^2}{2M_j} \rangle  = \mbox{Tr}\left[\frac{p_j^2}{2M_j} \;  \rho\right] 
\end{equation}
{\it and $\mathcal{E}_k(\omega)$ is given by Eq. (\ref{ho1}). The function $\mathbb{P}_j(\omega)$ is a probability density which obeys conditions  (\ref{poz}) and (\ref{nor}).} \\

\noindent The explicit form of the probability density $\mathbb{P}_j(\omega)$ is presented below. \\

{\it \bf Proof of Theorem 1: } To prove the relation (\ref{Ekj}), we apply the  fluctuation-dissipation relation of the Callen-Welton type \cite{call,kub}.   
One can exploit the results derived e.g.  in the Landau-Lifshitz book \cite{landau} [see Eq. (124.10)] or in the Zubarev book \cite{zubarev} [see Eq. (17.19g)]. We apply them  to 
the momentum operator $p_j$ of the system $S$. Without loss of generality we assume that  the average momentum $\langle p_j \rangle =0$ at the equilibrium state and then one obtains
\begin{equation}
	\label{landau}
	\langle p^2_j \rangle = \frac{\hbar}{\pi}\int_0^\infty  \coth{\left[\frac{\hbar\omega}{2k_BT}\right]} \,\chi_{jj}''(\omega)\, d\omega, 
\end{equation}
where $\chi_{jj}''(\omega)$ is the imaginary part of the generalized susceptibility,
\begin{equation}
	\label{zub2}
	\chi_{jj}(\omega) = \chi_{jj}'(\omega) + i\chi_{jj}''(\omega).
\end{equation}
The real part is an even function and the imaginary part is an odd function, 
\begin{equation}
	\label{real}
	\chi_{jj}'(\omega) = \chi_{jj}'(-\omega), \quad 	
	\chi_{jj}''(\omega) = - \chi_{jj}''(-\omega).
\end{equation}
The generalized susceptibility $\chi_{jj}(\omega)$ is the Fourier transform
\begin{equation} 	\label{green1}
	\chi_{jj}(\omega) = \int_{-\infty}^{\infty} \mbox{e}^{i\omega t} \, G_{jj}(t)\, dt
			\end{equation}
of the  response function $G_{jj}(t)$ which in fact  is the  retarded thermodynamic Green function \cite{zubarev}, namely, 
\begin{equation} 	\label{green2}
	G_{jj}(t)=  \frac{i}{\hbar} \theta(t) \langle [p_j(t), p_j(0)] \rangle,
\end{equation}
where $\theta(t)$ is the Heaviside step function and
\begin{equation} 	\label{green3}
p_j(t) = \exp(iH t/\hbar) p_j(0) \exp(-iH t/\hbar)
\end{equation}
is the Heisenberg representation of the momentum $p_j(0)$.
The averaging  in Eq. (\ref{green2}) is over the Gibbs canonical statistical operator (\ref{rho}). 

Now, we compare Eqs. (\ref{Ekj}) and (\ref{landau}), and obtain the  expression for 
 $\mathbb{P}_j(\omega)$ in the form 
\begin{equation} \label{chi}
\mathbb{P}_j(\omega) = \frac{2}{\pi M_j} \, \frac{\chi_{jj}''(\omega)}{\omega}.
\end{equation}
Hence, we obtain the formal expression for $\mathbb{P}_j(\omega)$. However, we have to show that it can be interpreted as  a probability density. \\

{\bf  Corollary}   {\it  The function  $\mathbb{P}_j(\omega)$  assumes  non-negative values for all positive values of the argument $\omega$. }\\

{\bf Proof:} We use the spectral representation of $\chi_{jj}''(\omega)$ in the form (see e.g. the equation just above Eq. (124.9) in the Landau-Lifshitz book \cite{landau}), 
\begin{equation} \label{posi}
\chi_{jj}''(\omega) = \frac{\pi}{\hbar} \; \left(1-\mbox{e}^{-\hbar \omega/k_B T}\right) 
\; \sum_{m,n} \rho_n |p_{nm}|^2 \delta(\hbar \omega + E_n -E_m), 
\end{equation}
where $p_{nm}$ are the matrix elements of the momentum operator in the basis of the eigenstates of the total Hamiltonian $H$, $E_n$ are the eigenvalues of $H$ and 
the population factor is $\rho_n = Z^{-1} \mbox{exp}(-E_n/k_B T)$. From  the form of this relation we see that for all positive $\omega$ the function $\chi_{jj}''(\omega)$ is positive and not zero. In fact, it is a well-known that  $\chi_{jj}''(\omega)$ is positive and is also named an absorptive part of the susceptibility 
$\chi_{jj}(\omega)$, see also the text below Eq. (123.11) in \cite{landau}. Hence, also $\mathbb{P}_j(\omega)$ given by Eq. (\ref{chi}) is positive for all positive values of $\omega$. \\

{\bf Theorem 2} {\it The function $\mathbb{P}_j(\omega)$ defined by Eq. (\ref{chi}) is normalized to unity, }
\begin{equation} \label{norm5}
\int_0^{\infty} \mathbb{P}_j(\omega) \,d\omega = \frac{1}{M_j} \; \frac{2}{\pi } \int_0^{\infty} \, \frac{\chi_{jj}''(\omega)}{\omega}\,d\omega =1.
\end{equation}
\\

{\bf Proof of Theorem 2:}  
According to the Kramers-Kronig dispersion relation
\begin{equation} \label{kramers1}
\chi'_{jj}(\omega) = \frac{2}{\pi} \;{\cal P} \int_0^{\infty} \, 
\frac{ u \chi''_{jj}(u)}{u^2 -\omega^2}  du,  
\end{equation}
where ${\cal P}$ denotes the principal value of the integral. Its value at $\omega =0$ 
reads 
\begin{equation} \label{kramers2}
\chi_{jj}(0) = \frac{2}{\pi} \; {\cal P} \int_0^{\infty} \, \frac{ \chi''_{jj}(u)}{u}  du,   
\end{equation}
where we utilize the relation $\chi'_{jj}(0)= \chi_{jj}(0) $ which follows from (\ref{zub2}) and (\ref{real}) for $\omega =0$.   The rhs of this equation is related to Eq. (\ref{norm5}). 
Alternatively, one  can apply  Eq. (123.19) in the Landau-Lifshitz book \cite{landau} which reads
\begin{equation} \label{kramers3}
\chi_{jj}(i\omega) = \frac{2}{\pi} \int_0^{\infty} \, \frac{ u \chi_{jj}''(u)}{\omega^2+u^2} \,du
\end{equation}
and for $\omega =0$ it takes the same value as (\ref{kramers2}). 
On the other hand, from Eqs. (\ref{green1}) and (\ref{green2}) it follows that
\begin{equation} \label{kramers5}
\chi_{jj}(0) =   \int_{-\infty}^{\infty} G_{jj}(t) \,dt = \frac{i}{\hbar} \; \int_0^{\infty}   \langle [p_j(t), p_j(0)]\rangle \,dt.
\end{equation}
We observe  that the problem of normalization of $\mathbb{P}_j(\omega)$ in Eq. (\ref{chi})
is converted to the problem whether the equality 
\begin{equation} \label{norm6}
\chi_{jj}(0) =  M_j 
\end{equation}
holds true for the  Hamiltonian (\ref{H})-(\ref{HSE}). 

This may seem surprising at first glance since $\chi_{jj}(0)$  does not  depend on the form of the potential, interaction, temperature and parameters of the Hamiltonian,  but it depends only on mass $M_j$ of the particle considered. 

In the next step, we prove that the relation (\ref{norm6}) indeed  holds true for a general form of the Hamiltonian (\ref{H}).  We start from the Heisenberg equations of motion for coordinate operators of the system $S$, namely,  
\begin{equation} \label{pp}
\frac{dx_j(t)}{dt} = \frac{i}{\hbar} \left[H, x_j(t)\right] = \frac{p_j(t)}{M_j}. 
\end{equation}
We insert it into Eq. (\ref{kramers5}) and obtain   
\begin{eqnarray} \label{kramers6}
\chi_{jj}(0) &=&    \frac{ i M_j}{\hbar}  \; \lim_{\epsilon \to 0^+} \int_0^{\infty} 
\mbox{e}^{-\epsilon t}  \frac{d}{dt} \langle [x_j(t), p_j(0)]\rangle \,dt \nonumber \\
&=&  \frac{ i M_j}{\hbar} \mbox{e}^{-\epsilon t}  \langle [x_j(t), p_j(0)]\rangle \vert ^{\infty}_0 
 +   \frac{ i M_j}{\hbar}  \lim_{\epsilon \to 0^+}  \epsilon \int_0^{\infty}  
\mbox{e}^{-\epsilon t}   \langle [x_j(t), p_j(0)]\rangle \,dt,   \nonumber\\
\ \ \
\end{eqnarray}
where we use a well-known limiting procedure with the $\epsilon$-term to ensure convergence of the integral \cite{kubo1}. 
The integral in the last line is finite and therefore this term tends to zero as $\epsilon \to 0$. In the first term,  for the  upper limit 
$t \to \infty$ the expression tends to zero. For the lower limit,  $\langle [x_j(0), p_j(0)]\rangle = i\hbar$.  
 Thus it finishes proofs of the relation (\ref{norm6}) and normalization of the function $\mathbb{P}_j(\omega)$ defined by Eq. (\ref{chi}).

\section{Comments and Discussion} 

\ \ \ 1.  The formula (\ref{Ekj}) is a generalization of the classical energy equipartition theorem. It fulfils  elementary  conditions for generalization:  
 Indeed, in the high temperature limit  

\begin{equation}
\coth\left[{\frac{\hbar \omega}{ 2k_BT}}\right] \approx \frac{2k_B T}{\hbar \omega}, 
\quad \mathcal{E}_k(\omega) \approx k_B T/2
\end{equation}
  and Eq. (\ref{Ekj}) reduces to its classical counterpart  
  \begin{equation}  \label{ET}
E_k^{(j)} = \frac{1}{2} k_B T \int_0^{\infty}  \mathbb{P}_j(\omega) \,d\omega =
  \frac{1}{2} k_B T
\end{equation}
because of normalization of $\mathbb{P}_j(\omega)$. We want to notice that Callen and Welton  in their 'historical' paper \cite{call}  missed the normalization: see Eq. (4.11) therein.  
 
 2. It has to be stressed that the formula (\ref{Ekj}) is universal, however, the mean kinetic energy $E_k^{(j)}$ depends  not only on temperature of the system (as in the classical case)  but also, via the probability density $\mathbb{P}_j(\omega)$,  on  a number of particles in the system, the form of the  potential  which acts on them, the form of interaction between particles and   strength of coupling between the system and thermostat. 

3.   If $H$ is the Hamiltonian of the composite system $S+B$ then all regimes, from weak to strong coupling with thermostat, can be analyzed.
However, if $H = H_S$ (there is no explicit interaction with thermostat $B$) then it means that only the weak coupling limit can be considered because  averaging is over the Gibbs canonical density operator $\rho_S \propto \mbox{exp}(-H_S/k_B T)$ valid in the weak coupling limit. 

4.  There are no specific assumptions regarding thermostat $B$: It should be infinitely extended and  satisfying the Kubo-Martin-Schwinger conditions expressing  periodicity of Green’s functions in imaginary time \cite{kubo1,martin}. 

5. The factor $\mathcal{E}_k(\omega)$ in Eq. (\ref{Ekj}) is the same as mean kinetic energy of a quantum harmonic oscillator in the Gibbs state $\rho_O \propto \mbox{exp}(-H_O/k_B T)$, where 
$H_O$ is the Hamiltonian of the harmonic oscillator \cite{feynman},  

\begin{equation}\label{ho}
\mathcal{E}_k(\omega) = 
\frac{1}{2m} \langle p^2 \rangle = 
\frac{\hbar \omega}{4} \coth{\frac{\hbar \omega}{ 2k_BT}}. 
\end{equation} 
It depends on the frequency $\omega$ of the harmonic oscillator but not upon its  mass $m$. However, in the considered model (\ref{H})-(\ref{HSE}), a harmonic oscillator does not occur at all.  It is a consequence of the above point 4 and the linear response theory \cite{ford17}. \\ 

6. As an example, we demonstrate how the above theory works for a free Brownian particle coupled to thermostat which is a collection of harmonic oscillators \cite{PRA}. What we need is the explicit form of the momentum operator $p(t)$ which has been calculated e.g. in Ref. \cite{PRA}, see Eq. (7) therein.   It reads 
\begin{eqnarray}\label{p}
p(t) = R(t)p(0) - \int_0^t  R(t-u) \gamma(u) \,du \, x(0) \nonumber \\
+ \int_0^t  R(t-u) \eta(u) \,du,
\end{eqnarray}
where $R(t)$ and $\gamma(t)$ are the response function and the memory kernel of the generalized Langevin equation. The operator $\eta(t)$ models quantum thermal noise and is expressed by thermostat operators which commute with the system operators. 
In Eq. (\ref{green2}), only the second term in r.h.s. of Eq. (\ref{p}) contributes to the commutator  yielding the Green function 
\begin{equation}\label{comut}
G(t) = \theta (t)  \int_0^t R(t-u) \gamma(u)\,du. 
\end{equation}
The susceptibility  $\chi(\omega)$ is a Fourier transform of the Green function $G(t)$ which is a convolution  in (\ref{comut}) of two scalar functions $R(t)$ and $\gamma(t)$.  Therefore as a result we obtain
\begin{equation}\label{chi2}
\chi(\omega) = {\hat R}_L(-i\omega) {\hat \gamma}_L(-i\omega),
\end{equation}
i.e., it is expressed by a product of two  Laplace transforms ${\hat R}_L(z)$ and  
 ${\hat \gamma}_L(z)$ of the functions $R(t)$ and  $\gamma(t)$, respectively. For the free Brownian particle of mass $M$ the Laplace transforms of  $R(t)$  reads \cite{PRA}
\begin{equation}\label{lapla}
{\hat R}_L(z) =  \frac{M }{M z  + {\hat \gamma}_L(z)} 
\end{equation}
and the  generalized susceptibility takes the form 
\begin{equation}\label{chifree}
\chi(\omega) = \frac{M {\hat \gamma}_L(-i\omega)}{-i\omega M + {\hat \gamma}_L(-i\omega)}.  
\end{equation}
It is seen that for any form of the memory function $\gamma(t)$ the value of susceptibility at zero frequency is the particle mass, $\chi(0)=M$. 

 In conclusion, applying the fluctuation-dissipation relation  we demonstrate  that Eq. (\ref{Ekj}) is valid for arbitrary quantum systems described by the Hamiltonian (\ref{H})-(\ref{HSE}) and being at the thermodynamic equilibrium state.   The probability distribution is of the form  (\ref{chi}),  where the susceptibility  $\chi_{jj}(\omega)$ is the Fourier transform of the  retarded thermodynamic Green function (\ref{green2}).  The formula (\ref{Ekj}) can be called the energy {\it partition} theorem for quantum systems because: (i) it is universal; (ii) it is  
 an extension of the formula for classical systems; (iii) it reduces to the energy {\it equipartition} theorem for high temperatures.

\section*{Acknowledgments}
 The author would like to thank P. H\"anggi and G.-L. Ingold for insightful  discussions on various aspects of this work and P. Talkner for suggestions regarding the proof of normalization.  The work  supported by the Grant NCN 2015/19/B/ST2/02856.




%
%

\end{document}